\documentclass{sigchi-ext}
\usepackage[T1]{fontenc}
\usepackage{textcomp}
\usepackage[scaled=.92]{helvet} 
\usepackage{graphicx} 
\usepackage{balance}  
\usepackage{booktabs} 
\usepackage{ccicons}  
\usepackage{ragged2e} 
\usepackage[caption=false]{subfig}

\def\plaintitle{SIGCHI Extended Abstracts Sample File: Note Initial
  Caps} 
\def\emptyauthor{}
\def\plainkeywords{Drone; Framework; Programming; Automation}

\copyrightinfo{This paper is published under the Creative Commons Attribution 4.0 International (CC-BY 4.0) license. Authors reserve their rights to disseminate the work on their personal and corporate Web sites with the appropriate attribution.\\
\emph{Interdisciplinary Workshop on Human-Drone Interaction (iHDI 2020)\\
CHI ’20 Extended Abstracts, 26 April 2020, Honolulu, HI, US}\\
\copyright~Creative Commons CC-BY 4.0 License.}

\title{Don't Drone Yourself in Work: Discussing DronOS as a Framework for Human-Drone Interaction}

\numberofauthors{5}
\author{
\alignauthor{
\parbox{\columnwidth}{\textbf{Matthias Hoppe, Yannick Weiß, \\Marinus Burger, Thomas Kosch}}\\
}
\alignauthor{}
\alignauthor{}
  \alignauthor{
\vspace{1mm} 
    \parbox{\columnwidth}{\affaddr{LMU Munich, Munich, Germany}}\\
      \parbox{\columnwidth}{\email{\{firstname.lastname\}@ifi.lmu.de}}\\
  }
}

\definecolor{linkColor}{RGB}{6,125,233}
\hypersetup{%
  pdftitle={\plaintitle},
  pdfauthor={\emptyauthor},
  pdfkeywords={\plainkeywords},
  bookmarksnumbered,
  pdfstartview={FitH},
  colorlinks,
  citecolor=black,
  filecolor=black,
  linkcolor=black,
  urlcolor=linkColor,
  breaklinks=true,
}

\teaser{
\vspace{-52mm}
{%
\setlength{\fboxsep}{0pt}%
\setlength{\fboxrule}{1pt}%
\fbox{\includegraphics[width=\columnwidth]{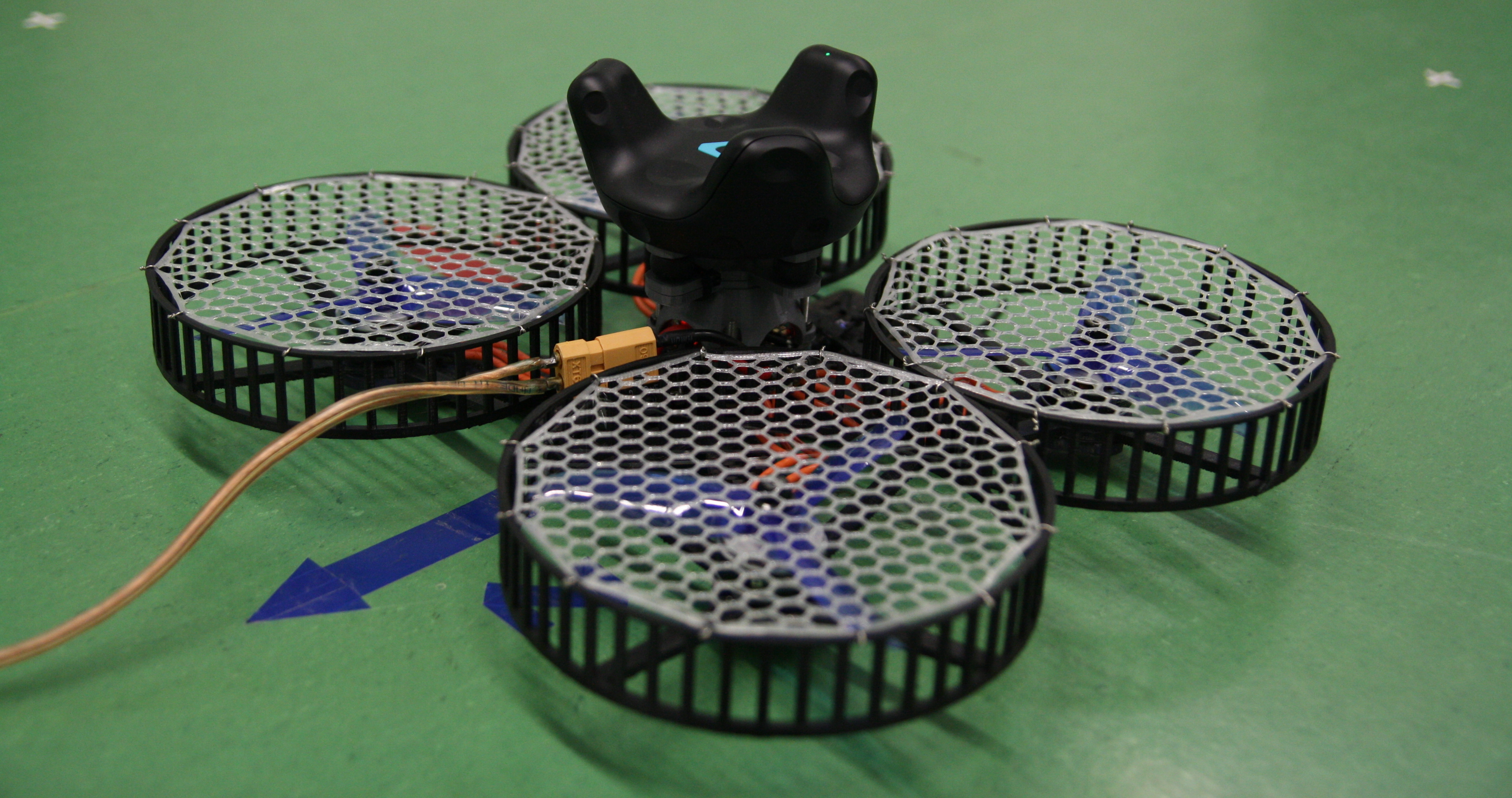}}%
}%
\caption{Custom drone build from of the shelf components. A mounted HTC Vive tracker enables low-cost tracking. We 3D-printed a rotor case for direct human-drone interaction.}
  \label{fig:teaser}
}

\begin{document}

\maketitle

\RaggedRight{} 

\begin{abstract}
More and more off-the-shelf drones provide frameworks that enable the programming of flight paths. These frameworks provide vendor-dependent programming and communication interfaces that are intended for flight path definitions. However, they are often limited to outdoor and GPS-based use only. A key disadvantage of such a solution is that they are complicated to use and require readjustments when changing the drone model. This is time-consuming since it requires redefining the flight path for the new framework. This workshop paper proposes additional features for DronOS, a community-driven framework that enables model-independent automatisation and programming of drones. We enhanced DronOS to include additional functions to account for the specific design constraints in human-drone-interaction. This paper provides a starting point for discussing the requirements involved in designing a drone system with other researchers within the human-drone interaction community. We envision DronOS as a community-driven framework that can be applied to generic drone models, hence enabling the automatisation for any commercially available drone. Our goal is to build DronOS as a software tool that can be easily used by researchers and practitioners to prototype novel drone-based systems.
\end{abstract}




\section{Introduction}
Drones are becoming commonplace for user interaction research~\cite{10.1145/3194317}. The increasing availability of consumer drones fosters the creation and prototyping of novel drone-based user interfaces. For example, drones can serve as a proxy for haptic feedback in virtual reality~\cite{10.1145/3282894.3282898, 10.1145/3027063.3050426}, navigation aid for persons with visual impairments~\cite{10.1145/2700648.2811362, 10.1145/3132525.3132556}, and as an observation unit for rescue operations~\cite{mayer:hal-02128385, 8551326}. Knierim et al.~\cite{10.1145/3173225.3173273} investigated a design space for different drone interaction modalities. Auda et al.~\cite{auda:proxydrone} investigated how drones can be used as a proxy for user interfaces. 

Some of these drones are delivered with vendor-dependent frameworks that enable flight automatisation through defined waypoints. These are usually programmed beforehand using defined targets or dynamic positioning relative to other objects. Most of these software frameworks are complicated or unstable to use and require prior programming knowledge. Furthermore, a plethora of options, such as the velocity of the drone, flight path corrections, or the surrounding has to be considered by the user. Contemporary available frameworks are often limited to a certain communication protocol, can only be applied on few drone models, and require manual programming of autonomous flight paths. To cope with this, Gomes et al.~\cite{10.1145/2858036.2858519} presented BitDrones, a toolbox that enables the programming of drones including interaction scenarios. The outlined interaction scenarios were sketched with custom-made drones that require users to reproduce them. Instead of using customised drones, Kosch et al.~\cite{10.1145/3266037.3266121} investigate the use of a remote controller for drone control. The remote control implements different visualisation modalities that communicate the target of the drone. Furthermore, several gestures were evaluated in the study. The use of an expensive tracking system remains as a key limitation of their work.

\begin{figure*}
   \centering
   \subfloat[][]{\includegraphics[height=0.65\columnwidth]{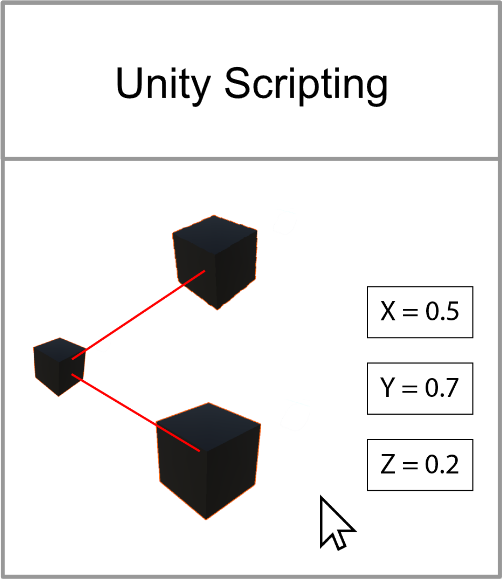} \label{lab:unityscripting}}
   \subfloat[][]{\includegraphics[height=0.65\columnwidth]{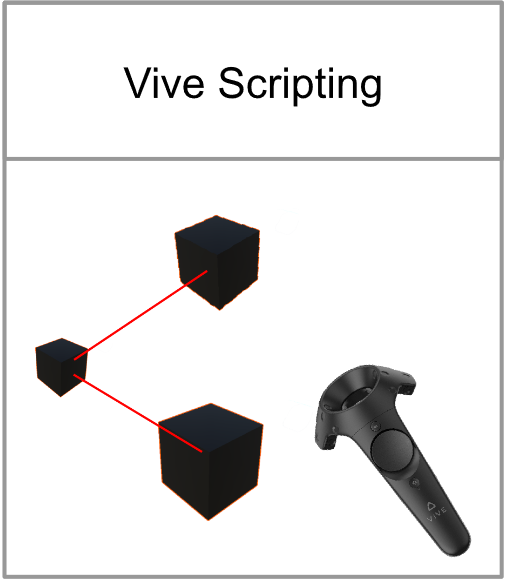} \label{lab:vivescripting}}
   \subfloat[][]{\includegraphics[height=0.65\columnwidth]{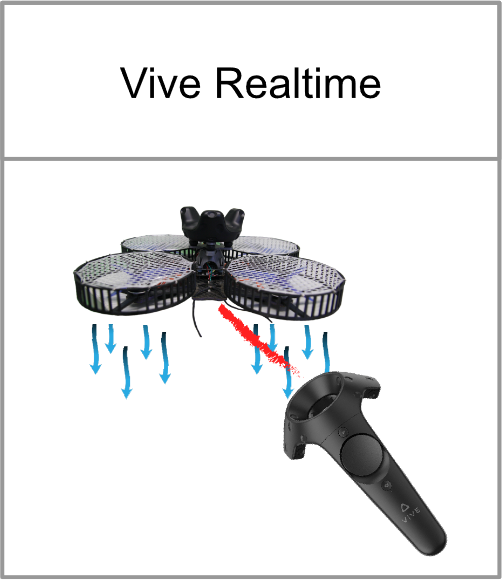} \label{lab:viverealtime}}
   \caption{DronOS supports three programming modes: (a) Unity Scripting for the advanced definition of waypoints, (b) Vive Scripting using point and click gestures, and (c) Vive Realtime where the drone follows the users' pointing direction.}
   \label{fig:sub1}
\end{figure*}

The implementation of interactive drone flight paths was subject to past research. However, the (a) use of proprietary communication protocols, (b) expensive tracking systems, (c) self-built drones, and (d) closed-libraries remain a key challenge of previous drone systems~\cite{hoppe:hal-02128388}. This workshop paper discusses DronOS, a framework that enables the interactive automatisation of drone flight paths using off-the-shelf components (see Figure~\ref{fig:teaser}). DronOS has been evaluated with three drone programming modalities in a previous user study~\cite{10.1145/3365610.3365642}, finding that users appreciated the usability of the system. Our overall aim is to establish DronOS as a community-driven framework for Human-Drone Interaction (HDI) researchers as well as practitioners. In the following, we elaborate on the basic concept of DronOS, explain the programming modalities, and present future research that we want to discuss with the HDI community.


\section{System Concept}
DronOS uses off-the-shelf hardware to implement the basic requirements for drone flight path definitions. DronOS employs HTC’s Lighthouse\footnote{\url{www.vive.com/eu/accessory/base-station}} tracking technology which is included in the HTC Vive kits. These offer a simple calibration procedure. The HTC Lighthouse system uses infrared to locate the position of an HTC Vive Tracker\footnote{\url{www.vive.com/eu/vive-tracker}}. We initially support this tracking system since it offers a low-budget tracking in contrast to, although professional, more expensive tracking systems. DronOS uses radio signals to communicate flight directions between a computer and a drone. Presets of PID controllers are available which can also be set manually for more experienced users. The drone itself is controlled via radio signals from a drone controller. This controller is connected to a computer that transmits the signal to the drone. DronOS supports three programming modes Unity Scripting, Vive Scripting, and Vive Realtime to define drone flight paths.

\begin{marginfigure}[-25em]
  \begin{minipage}{\marginparwidth}
   \centering
   \subfloat[][]{\includegraphics[width=0.9\marginparwidth]{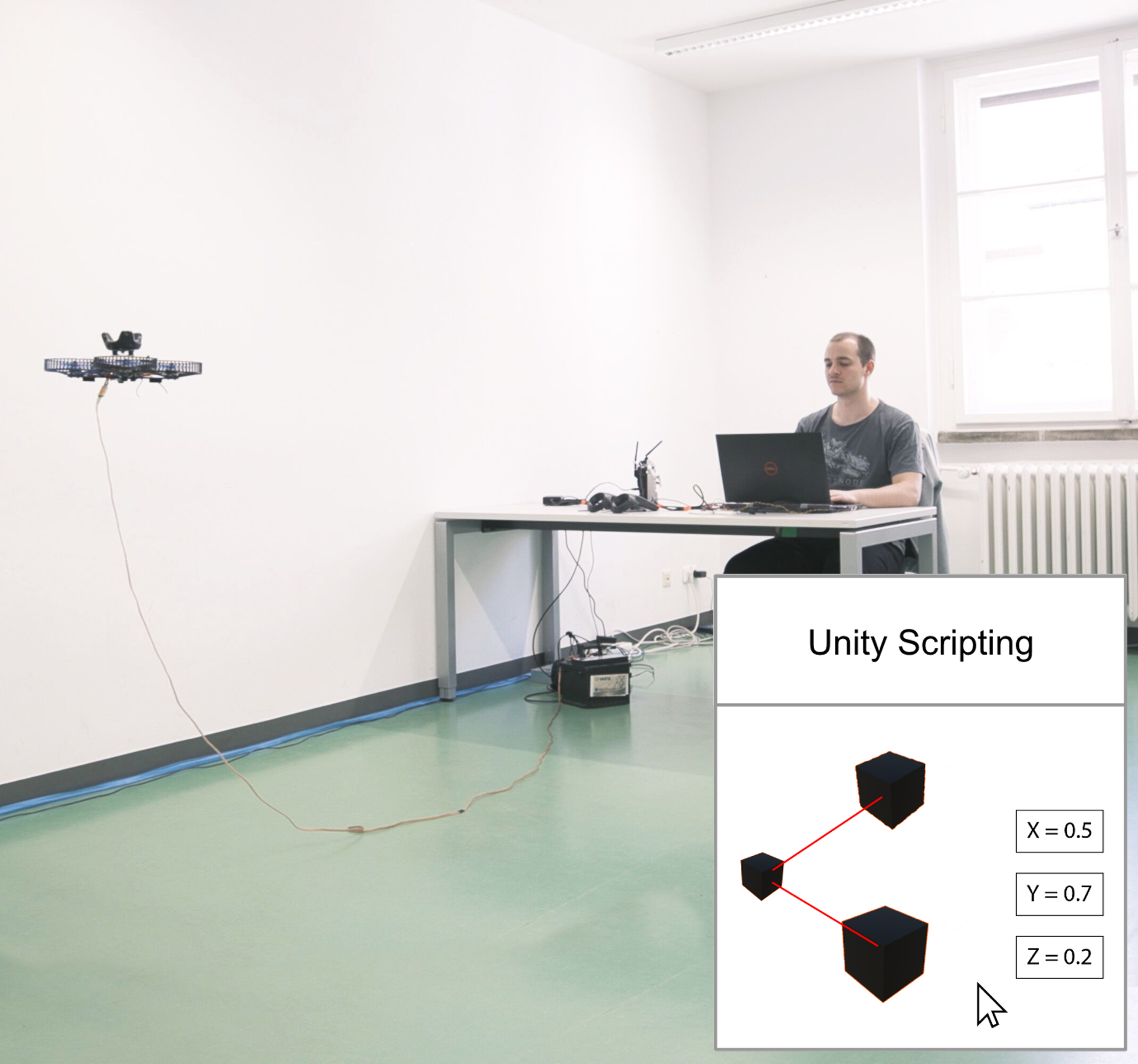} \label{lab:user_unityscripting}}\\
   \subfloat[][]{\includegraphics[width=0.9\marginparwidth]{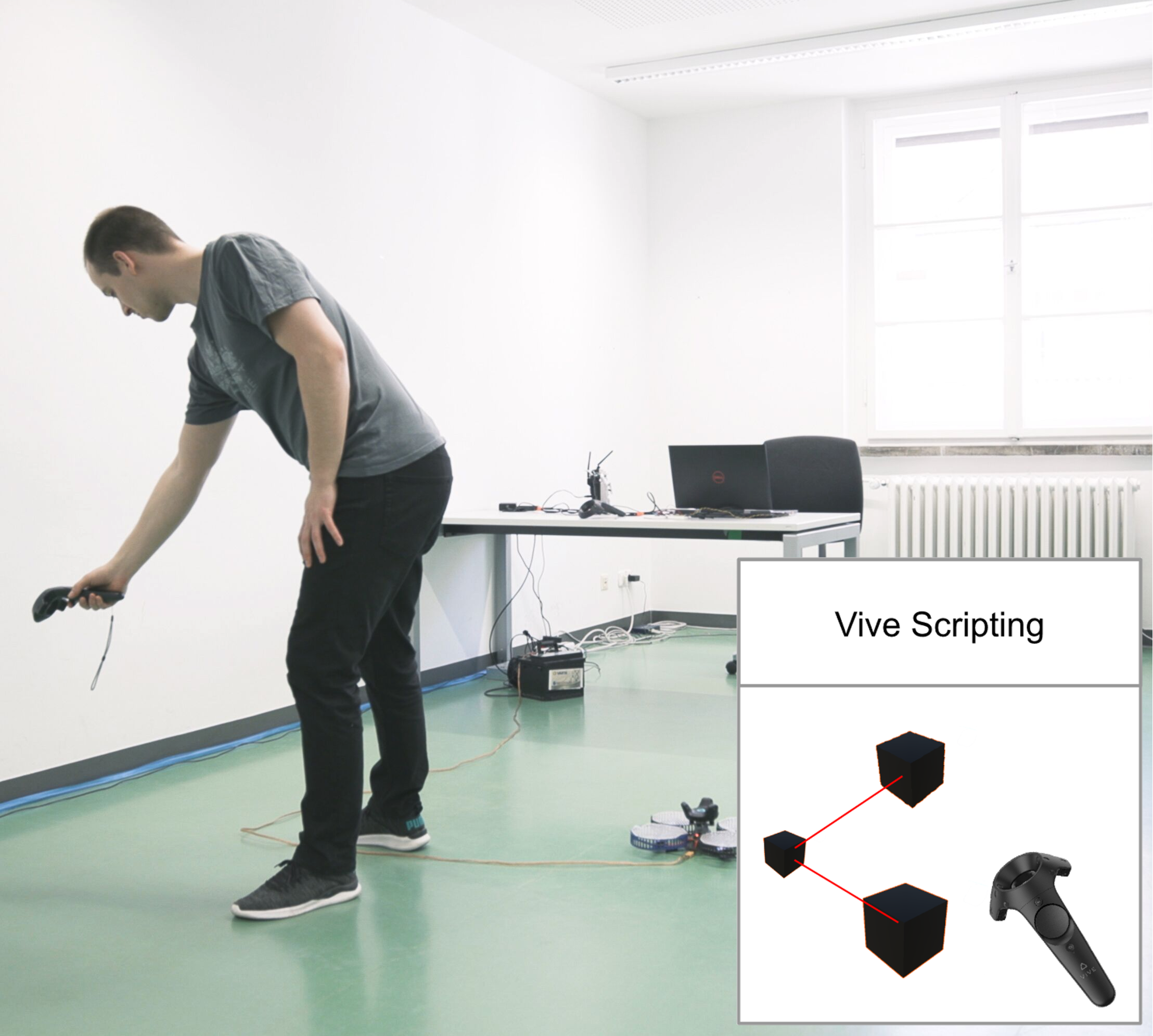} \label{lab:user_vivescripting}}\\
   \subfloat[][]{\includegraphics[width=0.9\marginparwidth]{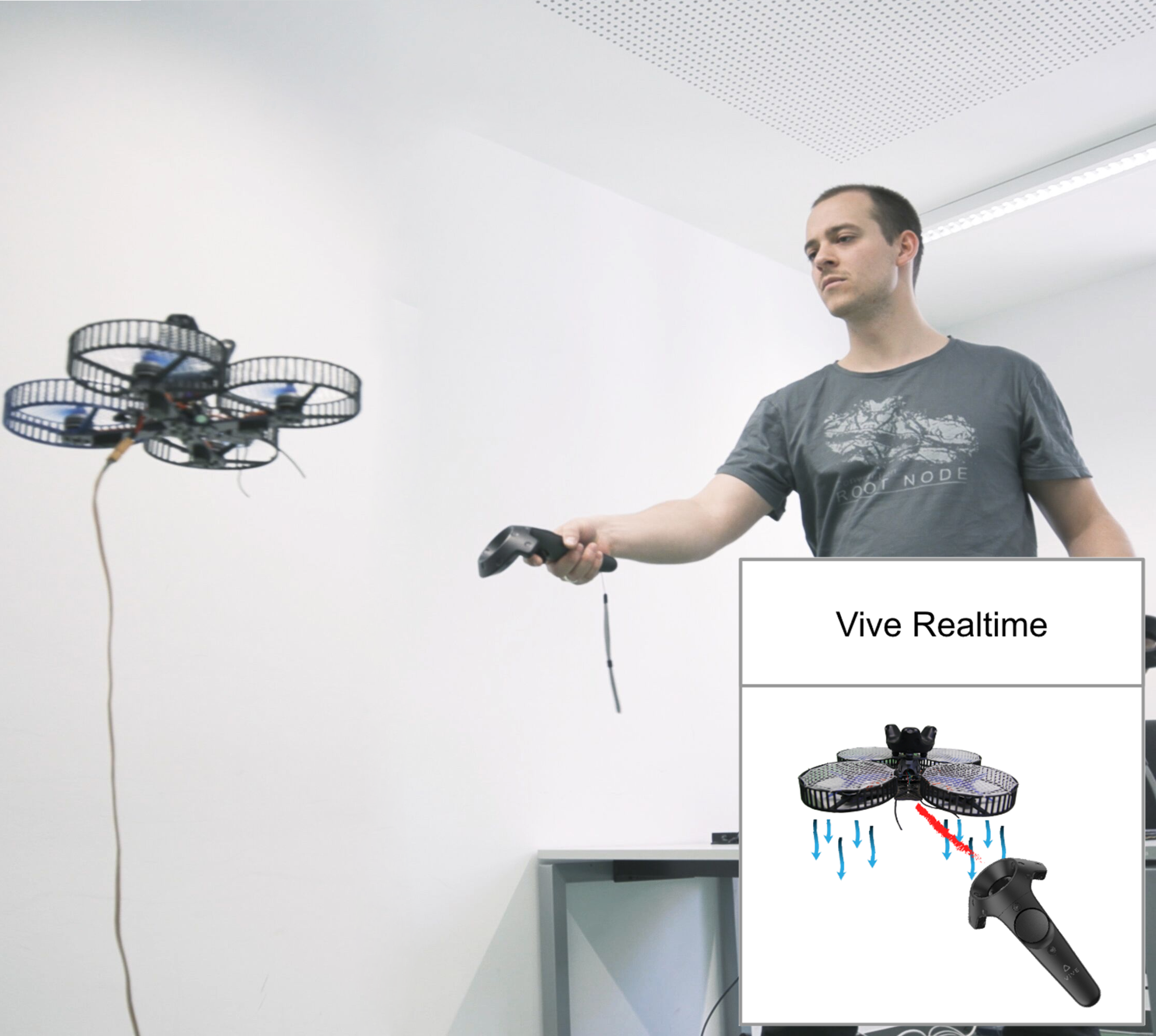} \label{lab:user_viverealtime}}\\
   \caption{User interacting with the modes (a) Unity Scripting, (b) Vive Scripting, and (c) Vive Realtime for creating flight paths.}
   \label{fig:sub1}
  \end{minipage}
\end{marginfigure}

\textbf{Unity Scripting} provides a user interface where flight paths can be defined using a Unity interface (see Figure \ref{lab:unityscripting} and \ref{lab:user_unityscripting}). New waypoints can be set using drag and drop. These are immediately visualised in 3D space and can be modified with advanced parameters.


\textbf{Vive Scripting}~
uses an HTC Vive controller to define waypoints in an ``programming by demonstration'' approach (see Figure \ref{lab:vivescripting} and \ref{lab:user_vivescripting}).
This allows the fast creation of flight paths without the need for graphical scripting or programming.

In \textbf{Vive Realtime} the drone levitates into the pointing direction of an HTC Vive Controller similar to the work of Kosch et al.~\cite{10.1145/3266037.3266121} (see Figure \ref{lab:viverealtime} and \ref{lab:user_viverealtime}). The distance between the controller and the drone can be adjusted. 

\section{Research Plan}
The current version of DronOS fully provides the aforementioned functionalities. DronOS includes the use of all self-built or off-the-shelf drones that support Betaflight. The framework enables users of all experience levels, from novice to expert, to create and redefine flight paths via both scripting and real-time control.

As human-drone interaction has special requirements, such as direct contact with the drone, we support researchers in the field of human-drone-interaction by implementing additional functions. Furthermore, DronOS is currently limited to the operation of one drone at a time. Hence, we plan to add functionalities that add the orchestration of multiple drones at the same time. This includes the communication between single drones to optimise the flying behaviour, such as avoiding collisions with other drones or users.

As safety is a key requirement when working with drones, we will include no-fly zones. These can be deployed as static areas (e.g. obstacles) where the drone will not be able to move and as dynamic areas (e.g., 20\,cm around a moving user) so that the drone reacts to the movement of the user. A core limitation is the use of an indoor tracking system. We explore alternative tracking modalities to realise omnipresent HDI use cases within the paradigm of ubiquitous computing. This includes the use of GPS and WiFi-based tracking that obviates the need for stationary tracking systems. Finally, we envision DronOS as a community-driven project. We continue to publish new features of the framework on Github to foster research and the implementation of new features within the HDI community\footnote{\url{www.github.com/hcum/dronos}}.

\section{Outlook}
This workshop paper presented DronOS, a generic framework that enables users to define the flight paths. We presented the currently available functionalities and operating principles. In contrast to the available features, we sketch a research plan with future features that will support researchers as well as practitioners in the development of future human-drone interfaces. We believe that our framework paves the way for the efficient deployment of drone interfaces.


\bibliographystyle{SIGCHI-Reference-Format}
\bibliography{extended-abstract}

\end{document}